\documentclass[prd,superscriptaddress,nofootinbib,11pt]{revtex4}
\usepackage{amssymb}
\usepackage{amsfonts}
\usepackage{amsmath}

\def\fs{\footnotesize}

\begin{document}

\title{From Dark Energy to Dark Matter via Non-Minimal Coupling}
\author{Andrzej Borowiec}
\email{borow@ift.uni.wroc.pl, borovec@theor.jinr.ru}
\affiliation{\fs Institute of Theoretical Physics, University of
Wroc{\l}aw\\
Pl. Maksa Borna 9, 50-204  Wroc{\l}aw, Poland.\\
and\\
Joint Institute for Nuclear Research, Dubna,
\\ Moscow region 141980, Russia}

\pacs{98.80.Jk, 04.20.-q}

\begin{abstract}
 Toy cosmological models based on non-minimal coupling between gravity and scalar dilaton-like field are presented
in the framework of Palatini formalism.
They have the following 
property: preceding to a given cosmological epoch is a dark energy epoch with an accelerated expansion.
The next (future) epoch becomes dominated by some kind of dark matter.
\end{abstract}

\maketitle

\subsection{\protect\bigskip Preliminaries and notation}
Modification of Eintein's General Relativity becomes viable candidate to address
accelerated expansion, dark matter and dark energy problems in modern cosmology
(see e.g. \cite{rev,Sergei} and references therein). This includes modified  theories with
non-trivial gravitational coupling \cite{GADED,DD3,Sergei2,Bertolami,Obukhov}.
Particularly, viable non-minimal models unifying
early-time inflation with late-time acceleration have been discussed in \cite{Sergei2}.

 Main object of our considerations in this note is cosmological applications of some non-minimally coupled
 scalar--tensor Lagrangians of the type
\begin{equation} \label{lag1}
L=\sqrt{g}\left( f(R)+F\left( R\right) L_{d}\right) +L_{mat}
\end{equation}
treated within  Palatini approach as in \cite{DD3}. Hereafter we set $L_{d}=-\frac{1}{2}g^{\mu \nu }\partial _{\mu }\phi \partial_{\nu }\phi$
the Lagrangian  for a scalar (massless) dilaton-like field $\phi$ and  $L_{mat}$  represents any matter Lagrangian.
Because of Palatini formalism  $R$ is a scalar  $R=R(g,\Gamma)=g^{\mu\nu}R_{\mu\nu}(\Gamma)$ composed of the metric $g$
and the Ricci tensor $R_{\mu\nu}(\Gamma)$ of the symmetric ($\equiv$ torsionless) connection $\Gamma$ ( for more details concerning Palatini formalism see e.g. \cite{FFV,ABF,Ricci2}).
Therefore $(g, \Gamma)$ are dynamical variables. Particularly, the metric $g$ will be used for raising and lowering indices.

We began with recalling  some general formulae  already developed in \cite{DD3}:
both  $f(R)$ and $F(R)$ are assumed to be analytical functions of $R$.
Dynamics of the system (\ref{lag1}) is controlled by the so-called master equation
\begin{equation}\label{master}
2f\left( R\right) -f^{\prime }\left( R\right) R+\tau =\left(
F^{\prime }\left( R\right) R-F\left( R\right) \right) L_{d}
\end{equation}
where prime denotes derivative with respect to $R$.
We set $T^{mat} =g^{\mu \nu }T _{\mu \nu }^{mat}$ and $ T^d=g^{\mu \nu
}T^d_{\mu\nu }=L_{d}$ for traces of the stress-energy tensors: matter $T^{mat} _{\mu\nu}=\frac{\delta L_{mat}}{\delta g_{\mu\nu}}$ and dilaton  $\ T^d_{\mu \nu }=-\frac{1}{2}\partial _{\mu }\phi $ $\partial _{\nu
}\phi $.

Equations of motion for gravitational fields $(\Gamma, g)$ can be recast \cite{DD3}
into the  form of generalized Einstein equations
\begin{equation} \label{hmeGEE4}
R_{\mu \nu }\left( h \right)\equiv  R_{\mu \nu }\left( bg\right)=g_{\mu \alpha }P_{\nu }^{\alpha }
\end{equation}
(see also \cite{ABF, Ricci2}), where
$R_{\mu \nu }(\Gamma)$ is now the Ricci tensor of the new conformally related metric $h=bg$. The conformal factor
$b$ is specified  below and a $(1,\,1)$ tensor $P_{\nu }^{\mu }$ is defined by:
\[
P_{\nu }^{\mu }=\frac{c}{b} \delta _{\nu }^{\mu }-\frac{F\left(
R\right) }{b}T_{\ \ \ \nu }^{d\ \mu }+\frac{1}{b} T^{mat\ \mu} _{\ \ \ \ \ \nu }
\]
Here one respectively has:
\begin{equation}\label{bc}
\begin{cases}
c=\frac{1}{2}\left( f\left( R\right) +F\left( R\right) L_{d}\right)
=(L-L_{mat})/2\sqrt{g} \cr
b=f^{\prime }\left( R\right) +F^{\prime }\left( R\right) L_{d}
\end{cases}
\end{equation}
Field equations for the scalar (dilaton-like) field $\phi$ is
\begin{equation}\label{dynd}
\partial _{\nu }\left( \sqrt{g}F(R)g^{\mu \nu }\partial _{\mu }\phi
\right) = 0
\end{equation}
which reproduces the same field equations as treated in \cite{DD3}.

\subsection{\protect\bigskip Cosmology from the generalized Einstein
equations}

We assume the physical metric $g$ to be a standard Friedmann-Robertson-Walker (FRW) metric $g$:
\begin{equation} \label{FRW}
g=-d t^2+a^2 (t) \left(d x^2+d y^2+d z^2\right)
\end{equation}
where $a (t)$ is a scale factor. 
We also suppose the Cosmological Principle to hold.
The matter content $T^{mat}_{\mu\nu}$ of the universe is
described by a non-interacting mixture of perfect fluids.
We denote by $w_i$  the barotropic coefficients. Each species is represented by the stress-energy tensor
$T^{(i)}_{\mu\nu}=(\rho_i+p_i)\,u_\mu u_\nu+p_i g_{\mu\nu}$ satisfying a metric (with the Christoffel  connection of $g$) conservation equation $\nabla^{(g)\,\mu}T^{(i)}_{\mu\nu}=0$  (see \cite{Koivisto}). This gives rise to the standard
relations between pressure and density
(equation of state) $p_i=w_i \rho_i$ and $\rho_i =\eta_i a^{-3\left(1+w_i\right) }$.

It follows thence that the generalized Einstein equations lead to the generalized Friedmann equation under the form:
\begin{equation}\label{genH}
\left( \frac{\overset{.}{a}}{a}+\frac{\overset{.}{b}}{2b}\right) ^{2}=
\frac{F(R)L_d}{6b}\ +\frac{c}{3b}+\sum_i\frac{(1+3w_i)\eta_i}{6b}a^{-3\left( 1+w_i\right) }
\end{equation}
where $(1+3w_i)\eta_i a^{-3\left( 1+w_i\right)}$ represents a perfect fluid component with an equation of state (EoS)
parameter $w_i$.

Let us observe that for standard cosmological model based on the standard Einstein-Hilbert variational principle
\begin{equation} \label{E-H}
L_{EH}=\sqrt{g}R+L_{mat}
\end{equation}
(considered both in purely metric as well as in Palatini formalism) the corresponding Friedmann equation
takes the form
\begin{equation}\label{sfe}
H^2\equiv \frac{\overset{.}{a}}{a}=
\frac{1}{3}\sum_i\eta_i\,a^{-3\left( 1+w_i\right) }
\end{equation}
when coupled to (non-interacting) multi-component perfect fluid.
This is due to the fact that geometry contributes to the r.h.s.  of the Friedmann equation through
$$
R=-T^{mat}=\sum_i\frac{(1-3w_i)\eta_i}{6b}a^{-3\left( 1+w_i\right) }
$$
For example, the preferred $\Lambda$CDM model requires three fluid components: cosmological constant $w_\Lambda=-1$, dust $w_{dust}=0$ and radiation $w_{rad}={1\over 3}$ and can be obtained from (\ref{genH}) provided $\alpha=\beta=\gamma=0$.

On the other hand we have that the field equation for the scalar field $\phi\equiv \phi(t)$ is
$ \frac{d}{dt}(\sqrt{g}F(R)\overset{.}{\phi})=0$,
so that  $\sqrt{g}F(R)\overset{.}{\phi }=const$ and consequently \ \ $gF(R)^{2}L_{d}=A ^{2}=const$.
This simply implies that:
\begin{equation}
\ F(R)^{2}L_{d}=A ^{2}a^{-6} \label{tor}
\end{equation}
with an arbitrary positive integration constant $A ^{2}$ (see (\ref{dynd})). \\

\subsection{\protect\bigskip Toy cosmological models}
Our objective here is to investigate a possible cosmological applications of the following subclass of gravitational Lagrangians (\ref{lag1})
\begin{equation}\label{lagr}
L=\sqrt{g}\left( R+\alpha R^2+\beta
R^{1+\delta}+\gamma R^{1+2\delta}L_{d}\right) +L_{mat}\end{equation}
where $\alpha, \beta, \gamma, \delta, $ are free parameters of the theory.  It should to be observed that the gravitational
part $f(R)$ contains Starobinsky term \cite{Starobinsky} with some $R^{1+\delta}$ contribution.
In the limit $\alpha\, ,\beta\, ,\gamma\rightarrow 0$ our Lagrangians reproduce General Relativity.
The numerical value for the constant $\gamma$  (when  $\neq0$) is
unessential since it can be always incorporated (by re-scaling) into  the field $\phi$.
As matter contribution we assume two non-interacting most natural components: pressureless  dust ($w_{dust}=0$) and radiation $w_{rad}={1\over 3}$.

Following common strategy particularly applicable within Palatini formalism  (see \cite{FFV,ABF,Ricci2}) 
one firstly finds out an exact solution of the master equation (\ref{master}). In the case under consideration it can be
chosen as
\begin{equation}
R=\left[\frac{\eta }{\left( 1-\delta\right) \beta }\right]^{\frac{1}{1+\delta}
}a^{-\frac{3}{1+\delta}}\equiv \xi a^{-\frac{3 }{1+\delta}}
\end{equation}
where $\xi\equiv\left[\frac{\eta }{\left( 1-\delta\right)\beta }\right]^{\frac{1}{1+\delta}}$
provided the the integration constant $A$ (see \ref{tor}) takes the value
\begin{equation}
A ^{2}=\frac{\gamma}{2\delta}\left[\frac{\eta}{\left(
1-\delta\right)\beta}\right]^{2}
\end{equation}
which can vanish only in the case $\gamma=0$ (no dilaton) and/or $\eta=0$ (no matter).

Then the conformal factor $b$ reads:
\begin{equation}\label{ab}
b
=\frac{1+4\delta}{2\delta}+2\alpha \xi a^{-\frac{3}{1+\delta}}+\beta (1+\delta) \xi^\delta a^{-\frac{3\delta}{1+\delta}}
\end{equation}

As a consequence, we have obtained the generalized Friedmann equations under the form:
\begin{eqnarray}\label{genFr}
\left( \frac{\overset{.}{a}}{a}+\frac{\overset{.}{b}}{2b}\right) ^{2}
=\left( 6b\right)^{-1}\Big[
\frac{1+\delta}{\delta}
\xi a^{-\frac{3}{1+\delta}}\ +\alpha\xi^2 a^{-\frac{6}{1+\delta}}\ +\cr
+\frac{\delta}{1-\delta}\eta a^{-3}+2\eta_{rad}a^{-4}\Big]
\end{eqnarray}

We are  now in position to calculate the generalized Hubble factor as
\begin{equation}
\frac{\overset{.}{a}}{a}+\frac{\overset{.}{b}}{2b}\equiv\frac{\overset{.}{a}}{a}\,B\equiv H\,B
\end{equation}
where $H$ denotes ordinary Hubble "constant" and
\begin{equation}
B=\frac{\frac{1+4\delta}{\delta}-2\frac{1-2\delta}{1+\delta}\alpha\xi a^{-\frac{3}{1+\delta}}+%
(2-\delta)\beta\xi^\delta a^{-\frac{3\delta}{1+\delta}}}%
{\frac{1+4\delta}{\delta}+4\alpha\xi a^{-\frac{3}{1+\delta}}+2\beta(1+\delta)\xi^\delta a^{-\frac{3\delta}{1+\delta}}}
\end{equation}
Before proceeding further let us observe that scaling properties of (\ref{genH})
\begin{eqnarray}\label{genFr2}
H^2\,B^2 =\left( 6b\right)^{-1}\Big[
\frac{1+\delta}{\delta}
\xi a^{-\frac{3}{1+\delta}}\ +\alpha\xi^2 a^{-\frac{6}{1+\delta}}\
+\frac{\delta}{1-\delta}\eta a^{-3}+\cr+2\eta_{rad}a^{-4}\Big]
\end{eqnarray}
are analogical to that in standard cosmology (\ref{sfe}). More exactly, choosing some reference epoch
$a_e\equiv a(t_e)$, e.g. the current cosmological epoch,  one can rewrite the generalized Friedmann
equation (\ref{genFr2}) as
\begin{eqnarray}\label{genFr3}
H^2\,B^2 =\left( 6b\right)^{-1}\Big[
\frac{1+\delta}{\delta}
\xi_e \left[\frac{a}{a_e}\right]^{-\frac{3}{1+\delta}}\ +\alpha\xi_e^2 \left[\frac{a}{a_e}\right]^{-\frac{6}{1+\delta}}\
+\frac{\delta}{1-\delta}\eta_e \left[\frac{a}{a_e}\right]^{-3}+\cr+2\eta_{rad, e}\left[\frac{a}{a_e}\right]^{-4}\Big]
\end{eqnarray}
where one has $\eta_e=\eta a_e^3$, $\eta_{rad, e}=\eta_{rad}a_e^4$, $\xi_e=\left[\frac{\eta_e }{\left( 1-\delta\right)\beta }\right]^{\frac{1}{1+\delta}}$,
$$
b=\frac{1+4\delta}{2\delta}+2\alpha \xi_e \left[\frac{a}{a_e}\right]^{-\frac{3}{1+\delta}}+\beta (1+\delta) \xi_e^\delta \left[\frac{a}{a_e}\right]^{-\frac{3\delta}{1+\delta}}
$$
etc..

Assume now that $0<\delta<1$.

Thus for $\frac{a}{a_e}\gg 1$,
one can approximate (\ref{genFr3}) by the following Hubble law
\begin{eqnarray}\label{genFr4}
H^2\,\approx{1\over 3}\Big[
\frac{1+\delta}{1+4\delta}
\xi_e \left[\frac{a}{a_e}\right]^{-\frac{3}{1+\delta}}\ +\frac{\delta \alpha}{1+4\delta}\xi_e^2 \left[\frac{a}{a_e}\right]^{-\frac{6}{1+\delta}}\ +\cr
+\frac{\delta^2}{(1+4\delta)(1-\delta)}\eta_e \left[\frac{a}{a_e}\right]^{-3}+\frac{2\delta}{1+4\delta}\eta_{rad, e}\left[\frac{a}{a_e}\right]^{-4}\Big]
\end{eqnarray}
Due to the factor $\delta$ the "true matter" and radiation decay as $\delta \mapsto 0$.  From the other hand the first
term on the r.h.s of (\ref{genFr4}) plays a role of matter: it can be considered as "dark matter" which amount is controlled by the factor $\xi_e$.
In the regime $\delta\mapsto 1$ the first term gives a bit of acceleration expansion while the second mimics matter (dark matter).

For $\frac{a}{a_e}\ll 1$ (preceding epoch), Friedmann type approximation reads instead
\begin{eqnarray}\label{genFr5}
H^2\,\approx{1\over 3}\Big[
\frac{(1+\delta)^3}{\delta(1-2\delta)^2\alpha}
 +\frac{(1+\delta)^2}{(1-2\delta)^2}\xi_e \left[\frac{a}{a_e}\right]^{-\frac{3}{1+\delta}}\ +\cr
+\frac{\delta(1+\delta)^2}{((1-2\delta)^2(1-\delta)}\frac{\eta_e}{\alpha\xi_e} \left[\frac{a}{a_e}\right]^{-\frac{3\delta}{1+\delta}}+\frac{2(1+\delta)^2}{(1-2\delta)^2}\frac{\eta_{rad, e}}{\alpha\xi_e}\left[\frac{a}{a_e}\right]^{-\frac{1+4\delta}{1+\delta}}\Big]
\end{eqnarray}
This epoch is dominated by  dark energy in the form of cosmological constant which produces Starobinsky inflation. The universe described by Freedmann equation (\ref{genFr5}) undergoes two additional phases of
accelerated expansion (power--law inflation) followed by the matter dominated era when $\delta\mapsto 0$. Similarly, when
$\delta\mapsto 1$. In this scenario the evolution goes from dark energy to dark matter dominated eras. More detailed study of such models will be given elsewhere.


\section*{Acknowledgment}
This note is dedicated to Sergei Odintsov on the occasion of his
birthday. I thank Gianluca Allemandi, Salvatore Capozziello and Mauro
Francaviglia for helpful discussions.

\end{document}